\documentstyle[12pt,aasms4]{article}
\begin{document}
\title{Luminosity Functions from
Photometric Redshifts I: Techniques}
\author{M. U. SubbaRao, A. J. Connolly and A. S. Szalay} \affil{The
Johns Hopkins University, Department of Physics and
Astronomy, The Johns Hopkins University, Baltimore, MD 21218}
\author{D. C. Koo} 
\affil{University of California Observatories, Lick Observatory,
Board of Studies in Astronomy and Astrophysics, University of
California, Santa Cruz, CA 95064}

\begin{abstract} The determination of the galaxy luminosity function
  is an active and fundamental field in observational cosmology. In
  this paper we propose a cost effective way of measuring galaxy
  luminosity functions at faint magnitudes. Our technique employs the
  use of galaxy redshifts estimated from their multicolor photometry
  (Connolly et al. 1995).  Associated with the redshift estimate is a well
  defined error distribution. We have derived a variant of
  \cite{Cmeth}'s (1971) C--method that considers, for each galaxy, the
  probability distribution in absolute magnitude resultant from the
  redshift error.  This technique is tested through simulations and
  potential biases are quantified.  We then apply the technique to a
  sample of galaxies with multicolor photometric data at moderately
  faint ($B \approx 23$) limits, and compare the results to a subset
  of these data with spectroscopic redshifts.  We find that the
  luminosity function derived from the photometric redshifts is
  consistent with that determined from spectroscopic redshifts.
\end{abstract}

\section{Introduction}
The determination of the galaxy luminosity function is of crucial
importance to cosmology.  Knowledge of the local luminosity function
is necessary to interpret the results of faint galaxy counts and their
implications for cosmography and galaxy evolution. A more explicit
measure of galaxy evolution is the variation of the luminosity
function with redshift and spectral type. Considering that the luminosity
functions may vary with redshift, color, local density, morphology,
spectral type, size, etc., many redshifts are required to get a
detailed picture of the multivariate galaxy distribution.

The difficulty in determining the luminosity function from a magnitude
limited sample of galaxies is that intrinsically bright galaxies can
be seen out to great distances while intrinsically faint galaxies can
be seen only relatively nearby.  One way to account for this
observational bias is to weight each galaxy's contribution by the
inverse of the volume over which it could have been observed.  This is
\cite{Vmax} $1/V_{max}$ technique.  The disadvantage of this approach
is that it does not properly account for galaxy clustering.
Nevertheless, it is still widely used today, (e.g.\ \cite{Lilly}
1995).  There are several non--parametric techniques which can account
for galaxy clustering: the method of \cite{cski}, the stepwise
maximum--likelihood method of \cite{EEP} (1988), and the C-method on
which our work is based (\cite{Cmeth} 1971).

Developments in the experimental determination of the luminosity
function have been tied to large redshift surveys.  Here we will
outline a few of the major contributions. The Stromlo-APM redshift
survey (\cite{Loveday} 1992) contained 1769 galaxies limited at
$b_J=18.7$ ($B\approx b_J + 0.3)$ and found a local luminosity
function which was well fitted by a Schecter function \cite{sch}, with
parameters $\phi^*=1.75\times 10^{-3}h_{50}^{-3}, M^*_{B_J}=-21.0 + 5
\log h_{50}$ and $\alpha=-1.0$.  The luminosity function from the CfA
redshift survey \cite{LGH89}, later extended in \cite{Marzke} (1994),
consists of 9063 galaxies at a limit of $m_z=17.0$.  Their luminosity
function had a much higher normalization $\phi^*=5.0\times
10^{-3}h_{50}^{-3}$ and three times as many faint galaxies as
predicted by the extrapolation of a flat ($\alpha=-1.0$) luminosity
function.  In contrast to these relatively local samples, the
Canada-France redshift survey (\cite{Lilly} 1995, henceforth CFRS) is
much deeper, containing 591 galaxies with a median redshift of
$\langle z \rangle=0.56$.  This landmark sample of galaxies allows
them to explicitly measure the evolution of the luminosity function
with redshift and even to separate the sample by color.

In this paper we describe a new technique which will allow the
determination of the luminosity function from deep samples with just
as many galaxies as the CFRS, but with far less observational cost.
The use of the photometric redshift technique of Connolly et al. (1995),
hereafter C95, provides
300 times as efficient use of telescope time as spectroscopic
techniques.  The tradeoff for the increased efficiency is an error in
the redshift determinations.  Our technique accounts for this error by
considering each galaxy as having a distribution in redshift.  This
distribution in redshift, folded in with the K--corrections, leads to a
distribution in absolute magnitude.  As pointed out by \cite{EEP} (1988) one
shortcoming of the standard non--parametric techniques for determining
the luminosity function is that they take no account of our
expectation that the luminosity function is smooth.  Our 
approach forces a degree of smoothness onto our luminosity function,
and in some sense the information lost due to the redshift errors is
partially compensated for by this smoothness assumption.

There are inherent difficulties in attempting to measure a continuous distribution such as the
luminosity function from discrete
galaxies.  The luminosity function techniques deal with this
difficulty either by binning the data, or in the case of the C-method
by producing a cumulative luminosity function with discrete jumps at
the location of each galaxy.  When the data are represented by
continuous functions such difficulties are naturally averted.  The
C-method is easily adapted to handle the continuous data, and the
calculation is actually simplified.

Throughout this paper we adopt the values $H_0=50 \ km \ s^{-1} \
Mpc^{-1}$ and $q_0=0.5$. Apparent and absolute magnitudes are denoted
by lower and upper case respectively.

\section{The Technique}
\subsection{Photometric redshifts}
Photometric redshifts were first proposed as an efficient means of
deriving statistical measures of galaxy distances by
\cite{Baum}. Early attempts to derive accurate estimates of galaxy
redshifts relied on fitting assumed galaxy evolution models to the
observed colors of galaxies \cite{Koo85}. This resulted in
uncertainties in the estimated redshifts in excess of 0.1 (for
redshifts greater than 0.35) and non--Gaussian error distributions.
More recently \cite{cz} have shown that by utilizing the flux
information and deriving empirical photometric-redshift relations from
existing deep redshift surveys, redshifts could be estimated with a
dispersion of $\sigma_z < 0.05$, out to $z=0.6$.

For this paper we assume that the error distribution in the estimated
photometric redshifts can be approximated by a Gaussian, consistent
with the findings of \cite{cz}. From the data of \cite{Munn}  we
find the dispersion in the estimated redshift, as a function of $r_F$
magnitude, to be well described by a linear relation,
\begin{equation}
\sigma_z = 0.051 + 0.0083(r_F-20).
\end{equation}

We use this relation to assign an error estimate for each galaxy both in
our simulations and in our application to observational data (see
3.1). We note that simulations used to determine the intrinsic dispersion of the photometric-redshift
relation have shown that the current estimate is dominated by the
photometric uncertainties in the \cite{Munn} data set. To determine
the applicability of our technique to future photometric samples, we
additionally simulate the effect of reducing the dispersion in the
photometric-redshift relation by a factor of three (see \cite{cz}).

\subsection{The `discrete' C-method} 
The C-method (c.f. \cite{Cmeth} 1971, \cite{Pet}) is a powerful estimator
for the luminosity function.  It is a non-parametric technique which
is insensitive to density inhomogeneities and makes full use of all the
data.  The C-method estimates the cumulative luminosity function which
then needs to be differenced in order to give the usual differential form.  In this
subsection we will describe the C-method and in the next we will
describe how we have extended it for use on a photometric redshift
data set.

We want to determine the cumulative luminosity function, $\Psi(M_0)$,
which is the density of galaxies whose absolute magnitude, $
M$ is brighter than, $M_0$ ($M<M_0$).  We actually measure the
cumulative distribution $X(M_0)$ which is $\Psi(M_0)$ subject to a set
of observational constraints (e.g.  apparent magnitude, surface
brightness limits). In general,
\begin{equation}
   {{d\Psi} \over {\Psi}}>{{dX} \over {X}},
\end{equation}
since it is generally harder to observe fainter galaxies.  We therefore wish to
construct $C(M)$, a subset of $X(M)$, for which the following
relationship holds:
\begin{equation}
   {{d\Psi} \over {\Psi}}={{dX} \over {C}}.
\end{equation}
To do this, we need to take the observational constraint on $dX$ and
apply that uniformly over $X$.  Namely, $C(M_0)$ is the number of
galaxies brighter than $M_0$ which could have been observed if their
absolute magnitude were $M_0$.  The quantities $C(M)$, $X(M)$, and
$\Psi(M)$ are illustrated in Fig 1.  To solve for
$\Psi(M)$, we integrate equation (3) i.e.,
\begin{equation}
    \Psi(M)= A \exp\left[\int\limits^M_{-\infty}{{dX} \over
{C}}\right].
\end{equation}
The quantity of interest, the differential luminosity function, is therefore,
\begin{equation}
    \Phi(M)= A \exp\left[\int\limits^M_{-\infty}{{dX} \over
 {C}}\right]{{dX(M)} \over {C(M)}} .
\end{equation}
When errors are not included $dX(M)$ is a series of Dirac delta
functions $dX(M)=\sum\limits_i \delta(M-M_i)$.  At the points $M=M_i$
the function $C(M)$ is indeterminate.  Because of this, the
determination of the integral in Eq. (4) requires further analysis. By
carefully considering the integral around these points, Lynden--Bell
arrives at the result
\begin{equation}
    \Psi(M)= A \exp\prod_i \left( {{1+C(M_i)}\over{C(M_i)}}\right).
\end{equation}

\subsection{Our modified, `continuous' C-method} 
In a photometric redshift survey we have a less accurate measure of
the galaxy's redshift.  The error distribution in redshift, folded in
with K-corrections, leads to a probability distribution in absolute
magnitude for each galaxy.  The function $dX(M)$ is now represented as a smooth function 
rather than a series 
of delta functions, and the integral in
Eqns (4) and (5) is easily calculated.  Each galaxy is represented as a
Gaussian distribution in redshift with mean $z_i$ and dispersion
$\sigma_i$.  The photometric redshift procedure is considered as a
Gaussian random process, and by integrating over these distributions we
recover the ``ensemble averaged'' luminosity function.  The
distribution in absolute magnitude includes the effects of
K--corrections and consequentially has a much more complicated form.
For this reason, when calculating the functions $C(M)$ and $X(M)$, we
prefer to work in redshift space. The following analysis is for a
complete magnitude limited sample.

Consider the function $z^*(m_i,M)$, which is the redshift at which the
$i$th galaxy will have apparent magnitude $m_i$ and absolute magnitude
$M$. It is the solution to the equation
\begin{equation}
 m_i-M = 5\log_{10}(d_l(z^*)) + 25 + K_i(z^*),
\end{equation}
where $d_l$ is the luminosity distance in Mpc, and $K_i(z^*)$ is the
K-correction for the $i$th galaxy at redshift $z^*$.  The function
$X(M)$ will include the fraction of the galaxy with $z>z^*(m_i,M)$.
For the case of a Gaussian error distribution this becomes
\begin{equation}
  X(M) = 0.5 \sum\limits_i {\rm erfc}\left({{z^*(m_i,M)
-z_i}\over{\sigma_{i}}}\right),
\end{equation}
where ${\rm erfc(x)}$ denotes the complimentary error function.
A galaxy with absolute magnitude $M$ can only be seen at redshifts
less than $z^*(m_{lim},M)$. The function $C(M)$ will include the
fraction of each galaxy between $z^*(m_i,M)$ and $z^*(m_{lim},M)$.
\begin{equation}
  C(M) = 0.5 \sum\limits_i \left[ {\rm erfc}\left({{z^*(m_i,M)
-z_i}\over{\sigma_{i}}}\right)-{\rm erfc}\left({{z^*(m_{lim},M)
-z_i}\over{\sigma_{i}}}\right)\right] .
\end{equation}
The relevant quantities are illustrated in Fig 2.  Once we have the
 two functions $C(M)$ and $X(M)$ the calculation of the luminosity
 function from equation (4) is straightforward.

One of the problems in constructing the galaxy luminosity functions is that
it is
not necessarily possible to solve for both the intrinsic luminosity
distribution and the variations of the density with redshift.  Using a
technique such as $1/V_{max}$ (Schmidt 1968), which makes the assumption
of uniform density, the only requirement is that the data
span the range in absolute magnitude over which we are estimating the
luminosity function.  Since we wish to make no assumptions regarding
the density fluctuations, we must make the further requirement that the
data be strongly connected.  Consider a situation where there are two
groups of galaxies, one near and one far.  The galaxies in the near
group are all too faint to be seen if they were transported to the far
group.  In this situation there is no way to determine the relative density fluctuations
between the near and far groups.  Hence it is impossible to construct
the full luminosity function.  Such a situation is indicated when
$C(M)=0$.  When this happens, the luminosity function is split into two
halves, brighter and fainter than $M$, which cannot be normalized with
respect to one another.  With our technique, where each galaxy
occupies a range in absolute magnitudes, the chances that $C=0$ for a
real magnitude limited sample of galaxies are greatly reduced.

\subsection{Normalization of the luminosity function}

The procedure described above produces an unnormalized estimate of the
luminosity function.  We normalize by calculating the expected number
of galaxies in a certain range in absolute magnitude and redshift and
comparing that to the number observed.  In order to do this
calculation we need to explicitly calculate the variation of galaxy
density with redshift.  This calculation can be done exactly like that of the luminosity function, simply by replacing absolute magnitude, $M$, with redshift, $z$ (see Eqns 4 - 9). We now have
unnormalized estimates of the luminosity function, $\Phi(M)$, and the galaxy
density as a function of redshift,  $\rho(z) dV/dz$.
The normalized luminosity function will be denoted by $\hat{\Phi}$.
To normalize the luminosity function we need the quantity $\rho_0$,
the integral of the luminosity function.  We need to choose a range of
redshift$(z_0,z_1)$ over which to calculate $\rho_0$:
\begin{equation}
\rho_0 ={{\int\limits_{z0}^{z1}{ \rho(z){dV}\over {dz}}dz}
\over{{\int\limits_{z0}^{z1}{{dV}\over {dz}}dz}}}.
\end{equation}
We can now normalize by integrating the unnormalized
luminosity function and the relative density fluctuations,
$\rho(z)/\rho_0$ to get the expected number of galaxies.  Comparing
this number with the observed number of galaxies will give us the
normalization.  However, to do this integral we need to know the
intrinsic distribution of galaxy types and their K--corrections.  Therefore
we must limit the integral to a range of absolute
magnitude and redshift where the entire spectrum of galaxy types can
be observed.  We can now calculate the number of galaxies which we
expect to observe in the range $M0<M<M1$ and $z0<z<z1$.
\begin{equation}
N_e=\sum\limits_{i=1}^4f_i\int\limits_{M0}^{M1}\int\limits_{z0}^{z1}A\Phi(M)
{{\rho(z)}\over{\rho_0}}{{dV}\over{dz}}\Theta(z^*(m_{lim},M)-z)dz dM,
\end{equation}
where $f_i$ represents the fraction of galaxies assigned to the $i$th
spectral type based on their $b_J-r_F$ colors, and $\Theta(x)$ is a
step function.  The number of galaxies observed between the redshift
and absolute magnitude limits is $N_o$.  Since the galaxies are
represented as a distribution both in redshift and absolute magnitude
this is not a integer quantity.  Comparing these quantities gives us
the normalization,
\begin{equation}
\hat{\Phi}(M)=\Phi(M){{N_o}\over{N_e}}
\end{equation}

\subsection{Errors and Biases}
Errors are estimated using bootstrap techniques.  Artificial samples
are generated by randomly picking galaxies from the real sample.  Each
artificial sample has the same number of galaxies as the original
sample.  Individual galaxies may be picked multiple times or not at
all. 100 samples are generated in this manner and the luminosity
function is calculated for each.  The variance in the luminosity
function is then given by the variance of the artificial samples.  We
use the random number generator {\tt ran1} from Numerical Recipes
\cite{numrecs}.

It is well known that the C-method is an unbiased estimator of the
luminosity function.  However, when we are dealing with fuzzy data
this will no longer be the case.  The broad distribution in absolute
magnitude acts as a smoothing kernel, preferentially scattering
galaxies away from $M^*$, where the distribution is peaked, and
towards the bright and faint ends of the distribution.  This produces
a bias analogous to Eddington or Malmquist bias (\cite{Ed},
\cite{Malm}).

We have attempted a correction for this bias.  Since the absolute
magnitude distribution is different for each galaxy depending on its
redshift, redshift error and spectral type; the correction is best
handled by Monte-Carlo techniques.  The problem is as follows:
given a measured absolute magnitude distribution
$\left({{dX(M)}\over{dM}}\right)_m$, what is the most likely true
distribution $\left({{dX(M)}\over{dM}}\right)_t$.  Once this is known
the luminosity function can be corrected using,
\begin{equation}
\hat{\Phi}(M)=\Phi(M)\left({{dX(M)}\over{dM}}\right)_m^{-1}
\left({{dX(M)}\over{dM}}\right)_t.
\end{equation}
We can estimate the effect of going from the true to the measured
distribution by randomly scattering each galaxy in redshift according
to its error distribution.  We invert this procedure in an iterative
manner.  The noisiness of the sample prevents absolute convergence so
we have terminated the procedure after two iterations.  We have
performed simulations to estimate the effect of both the bias and its
correction.

Simulated galaxies are placed within the simulated volume according to
a given Schecter function.  Galaxy types are randomly assigned to one
of the four templates from \cite{Kinney} (1996), as described in
section 3.2.  The apparent magnitude is then calculated using the true
redshift and the galaxy's K-correction. The galaxy is either included
or excluded from the sample according to its apparent magnitude.  The
galaxy's $b_J-r_F$ color is determined from the template which is used
to assign the error in redshift as in Eqn 1.  The estimated redshift is
then scattered from the true redshift by an amount drawn from its
Gaussian error distribution.

We present the results from four simulation scenarios with 100
realizations each.  The number of galaxies in the simulations is
allowed to vary with each realization. The normalization of the input
luminosity function is chosen so that there are roughly 800 galaxies
in each of the simulations. We vary both the input luminosity function as
well as the error distribution.  As discussed in \cite{cz}, the errors
of Eqn 1 are dominated by errors in the photographic photometry.  With better
photometry galaxy redshifts could be estimated three times as
accurately.  Simulations are performed for both the present as well as
potential errors on both steep and flat luminosity functions
($\alpha=-1.5$ and $\alpha=-1.0$).

Fig 3 shows the deformation vectors for the recovered Schecter
parameters for the simulations.  In all cases the slope is
overestimated, although our correction does seem to reduce the
effect. We slightly over--correct the bright end, systematically
overestimating $M^*$ by 0.2 magnitudes. Fig 4 shows the recovered
luminosity functions with the input Schecter luminosity function.
Steep luminosity functions are reproduced much better than flat ones.
This is easily understood as there are many more galaxies at faint
magnitudes in the steep case.  When viewing the figures it is useful
to note that fainter than $B_J=-17.0$ the entire contribution to the
luminosity function is from galaxies with $z<0.1$, where the relative
errors in redshift are very high $\sigma_z/z \approx 1$.  The
uncertainty in deriving the luminosity function is dependent on the
redshift error, $\sigma_z$.  When we perform the simulations using the
errors expected from higher quality photometric data the luminosity
functions are recovered more accurately (see Fig 3 and Fig 4).

\section{Application of the Technique}

\subsection{The Data}
We construct a sample of galaxies with photometric redshifts from the
photometric and spectroscopic survey data of Koo and Kron
(\cite{Kron80}, \cite{Koo86}). These data consist of scans
of photographic plates taken with the KPNO 4m. They cover the $U$,
$B_J$, $R_F$ and $I_N$ passbands and are 50\% complete at a magnitude
limit of $B_J \sim 24$. Details of the photometric data can be found
in (\cite{mab} 1994, and references therein), while the spectroscopic data
are described in \cite{Munn}.

Our current analysis considers only the high Galactic latitude field
Selected Area 68 (SA68). In this paper we describe the technique, and we 
do not attempt to combine fields in order to minimize possible errors from
zero point offsets amongst different plate material.  From these data
we define two samples of galaxies. We derive a $b_J < 22.5$ magnitude
limited sample of galaxies detected in all four passbands from which
we can determine photometric redshifts. At this magnitude limit the
relative uncertainties in the photometric data were less than 0.5
magnitudes in each of the four passbands. From the photometric sample
we derive a subset of those galaxies with high quality spectroscopic
redshifts. We use these two data sets to compare the relative accuracy
and robustness of our analysis. The photometric redshift sample
consists of 772 galaxies and the spectroscopic redshift subset has 114
galaxies.

We use the techniques described in \cite{cz} to estimate the galaxy
redshifts from their multicolor broadband photometry. We apply the
fits from \cite{cz} of a second order polynomial to the four optical
passbands and derive a relation between spectroscopic redshift and
broadband photometry. The photometric redshift distribution of this
sample is given in Fig 5. Comparing the spectroscopic and photometric
redshifts we find that, at $b_J<22.5$, the dispersion in the
photometric redshift relation is Gaussian with $\sigma_z =0.045$
slowly increasing towards fainter magnitudes. The correlation between
the redshift dispersion and $r_F$ magnitude is given in Eqn 1.

\subsection{K-corrections}

We calculate the K-correction for each galaxy from the spectral energy
distributions (SEDs) of \cite{Kinney} . We choose these spectra
over stellar synthesis models as they are derived from actual galaxy
spectra. We selected five SEDs from the \cite{Kinney} data: those of an
Elliptical, S0, Sb and two starburst galaxies with E(B-V) = 0.05 (S1)
and E(B-V) = 0.70 (S6). These five SED's were chosen to encompass the
expected color distribution of our galaxies. The Elliptical and S0
templates have very similar colors, so they were averaged together, and the
remaining four templates were used in calculating the K-corrections.
Fig 6 shows the $b_J - r_F$ colors of these templates as a function of
redshift.

For each galaxy in the photometric and spectroscopic samples we derive
an estimate of its redshift. We determine the spectral type of a
galaxy from its observed $b_J - r_F$ color. To improve the accuracy of
our K-corrections we interpolate between the two galaxy templates with
the closest $b_J - r_F$ colors. From the estimated redshift and
spectral type of each galaxy we calculate the K-corrections in each of
the four bandpasses. The distribution of apparent $b_J - r_F$ color of
the photometric sample is shown in Fig 6.
 
\subsection{Completeness}
There are two sources of incompleteness that we need to be concerned about.
  The first is incompleteness in the photometric sample, and the
second arises from the estimation of redshifts.  We believe the first
source of incompleteness to be minimal. The sample was conservatively
cut at $b_J=22.5$, which is a full magnitude and a half below the 50\%
completeness level (Koo 1986).  The second source of incompleteness
has its origin in the fact that the redshift estimation procedure
occasionally indicates a negative redshift.  We interpret the negative
redshifts as reflecting the error distribution function which extends
below zero for the lowest redshift galaxies.  49 of the 772 galaxies
(6.3\%) had redshift estimates which were negative.  These galaxies 
are essentially lost from our sample. 
We consider three methods of dealing with their omission.  The first
is to simply to ignore them, this is the `minimal' luminosity
function.  This has the advantage that it makes no assumptions about the
missing data; however, in this case our luminosity function is
strictly a lower limit.  The second method is to weight the luminosity
function by the missing fraction.  This is the `weighted' luminosity
function.  This method makes the explicit assumption that the
distribution of the missing galaxies is identical to those whose
redshifts we could estimate.  This assumption is supported by the fact
that the incompleteness fraction appears to be a weak function of
apparent magnitude. If the incompleteness is only dependent on
redshift, then the incompleteness will leave our luminosity function
estimates unbiased.  Yet we know from Eqn 1 that this is not entirely
true. The third method takes advantage of our expectation that the
source of incompleteness is low redshift galaxies and cuts the low
redshift galaxies from the sample. This is the `cut' luminosity
function.  The `cut' luminosity function has the further advantage
that it is in the low redshift regime where the error distribution in
absolute magnitude is the broadest.  We cut from the sample the
contributions of galaxies with $z<0.1$.  Remember that in our scheme
galaxies occupy a distribution in redshift.  When we make a cut in
redshift we only consider the fraction of each galaxy's probability
distribution between the redshift limits.

\subsection{Results}
We have applied our luminosity function technique, including bias
corrections, on the photometric redshift sample described above.  We
derive luminosity functions for the `minimal', `weighted' and
`cut' galaxy samples. In Fig 7  the `minimal' luminosity
function is shown bracketed by lines denoting a one sigma standard deviation.
The luminosity function is represented as a curve rather than the
usual points since our technique produces a continuous estimate of the
luminosity function.  Fig 7 also shows the number of galaxies as a
function of absolute magnitude.  This shows roughly how many galaxies
are contributing to the luminosity function estimation in each
magnitude interval.

The luminosity functions for each of the three samples are reasonably
well fit by Schecter functions. Table 1 shows the best fit Schecter
parameters for the three luminosity functions. For all of the samples,
the parameters derived are consistent within one standard deviation.

The values of $M^*$ and $\Phi^*$ agree within the errors to the values
derived by \cite{Loveday} (1992).  Yet our luminosity functions appear to
have a steeper faint end slope, even when the bias in Fig 3 is taken
into account.  The luminosity function of the CfA redshift survey also
shows a considerable faint galaxy excess (\cite{Marzke} 1993). 
Keeping in mind that the photometric redshift estimation tends to
smooth the luminosity function, we
might expect a luminosity function with a strong faint end turn up, 
such as that of
the CfA, to be better fit by a Schecter function with a steeper faint
end slope. The large number of inferred faint galaxies  also helps
to reconcile the number counts and the galaxy redshift
distribution \cite{Koo93}.

We apply a further test of our technique by comparing the results from
the photometric sample to the luminosity function derived from a
sub-sample of 114 galaxies with high quality spectroscopic redshifts.
These galaxies are from one of a number of fields whose luminosity
function is estimated in a upcoming paper by \cite{kbs}.  In order to
fairly compare the spectroscopic sample with the deeper photometric
sample we have cut back the sample to $b_J=20.0$.  The results are
shown in Fig 8. The luminosity derived from the photometric data is shown
by a continuous curve, and that from the spectroscopic redshift
sample by the points. It is readily apparent that the two luminosity functions agree very well.

\section{Conclusions}
We have presented a new technique for the determination of luminosity
functions from photometric redshift samples.  This technique considers
the statistical scatter in the photometric redshift procedure and integrates
over the redshift probability distribution for each galaxy.  The result is a
continuous estimate of the luminosity function.  This procedure can
easily be generalized for use when the errors in any of the parameters are
large.  The large cost benefit in obtaining photometric
vs. spectroscopic redshifts indicates that this technique can be an
important tool in the multivariate analysis of galaxy properties.

In conclusion we would like to make the following points regarding our
technique and its application:

(1) The results from our technique are in good agreement with those
from deep spectroscopic surveys.  Our results do seem to indicate a
very large number of faint galaxies.  However, at the faint end of the
luminosity function there appears to be little agreement among
different groups (\cite{Loveday} 1992,\cite{Marzke} 1994,\cite{Lilly} 1995).

(2) Our technique will do better as we go to fainter samples and hence
higher redshift galaxies.  Low redshift galaxies are difficult to deal
with for two reasons.  The first is that the size of the error distribution
in absolute magnitude diverges as you approach $z=0$.  The second
reason is that the photometric redshift estimation procedure
occasionally indicates a negative value for some, presumably, low
redshift galaxies.  These galaxies create incompleteness in our
sample.  

(3) Our technique is currently the most practical method for pursuing a
multivariate study of the galaxy luminosity function.  It is certainly
informative to split the luminosity function in the two dimensions of
redshift and spectral type (\cite{Lilly} 1995).  The efficiency of the
photometric redshift estimation makes it conceivable to split the
luminosity function in three or even four dimensions.

\acknowledgements 
We thank Michael Fall and Matthew Bershady for
helpful discussions, Richard Kron for the use of the data
set, and Amanda Marlowe SubbaRao for help improving the readability of
the text. A.J.C. and A.S. acknowledge partial support from NSF grant
AST-9020380, an NSF-Hungary Exchange Grant, the US-Hungarian Fund and the
Seaver Foundation. D.C.K. acknowledges support from NSF grant AST-99-58203.

\newpage \figcaption[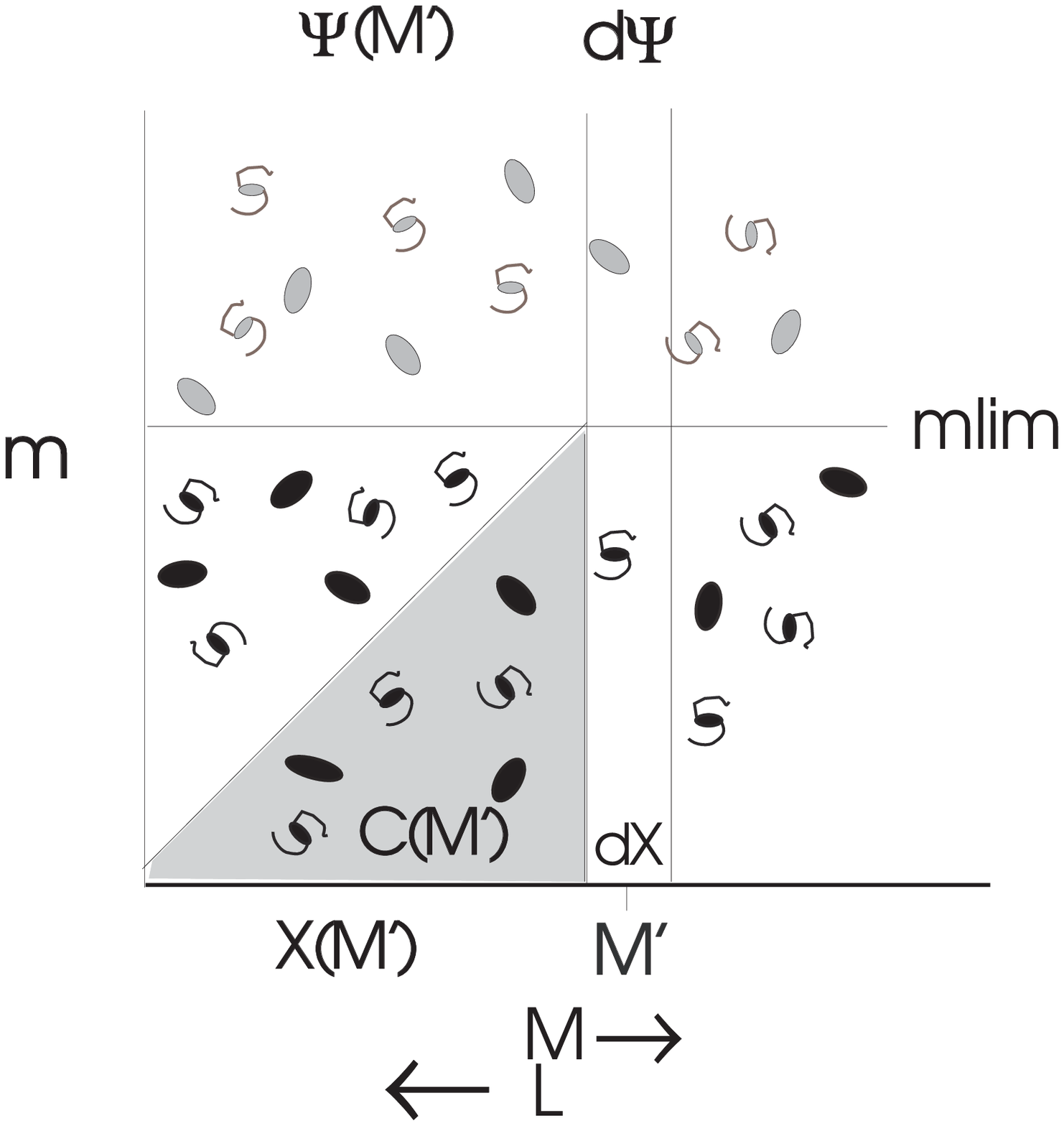]{ This figure illustrates the mathematical quantities defined in the derivation of the C-method.  We are only able to
observe the galaxies below the limiting magnitude line.  In this apparent vs.
absolute magnitude plot, a 45 degree line
represents, neglecting differential K-corrections, a
line of constant distance.
$\Psi$ is the true number of
galaxies brighter than M. X(M) is the observed number of galaxies
brighter than M.  In general $ {{d\Psi} \over {\Psi}}>{{dX} \over
{X}}$ since it is generally harder to observe fainter galaxies.  We define the quantity C so
that ${{d\Psi} \over {\Psi}}={{dX} \over {C}}$. C is the number of
galaxies in the shaded region of the figure.  Each galaxy in this region could have been observed if their absolute magnitude was $M'$.}

\figcaption[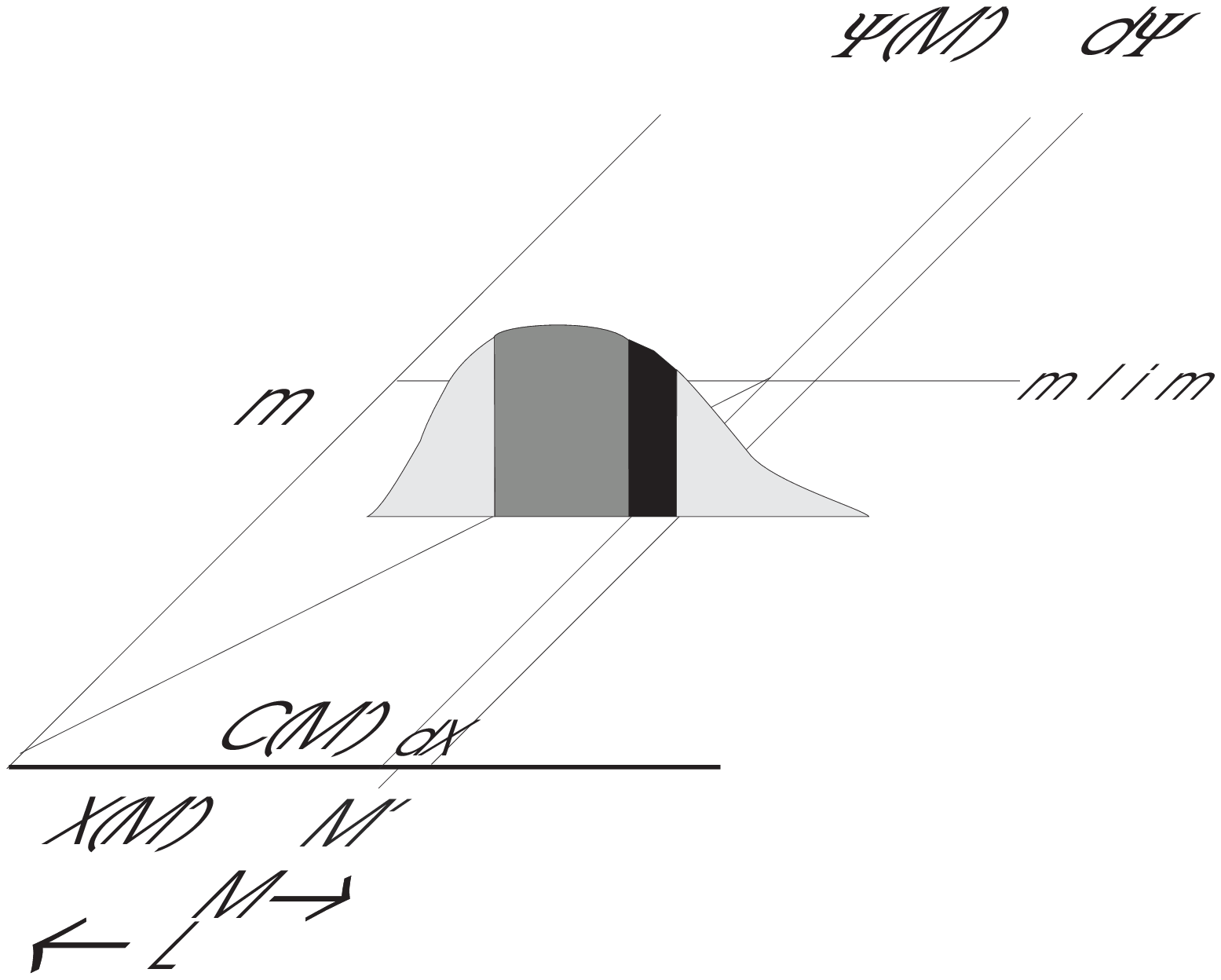]{This figure illustrates the the same quantities as Fig 1, but now each galaxy is represented as a probability distribution in absolute magnitude. This probability distribution results from the error associated with the photometric redshift estimation.  The fraction of a galaxy
which contributes to the quantity C is the middle shade of grey.}

\figcaption[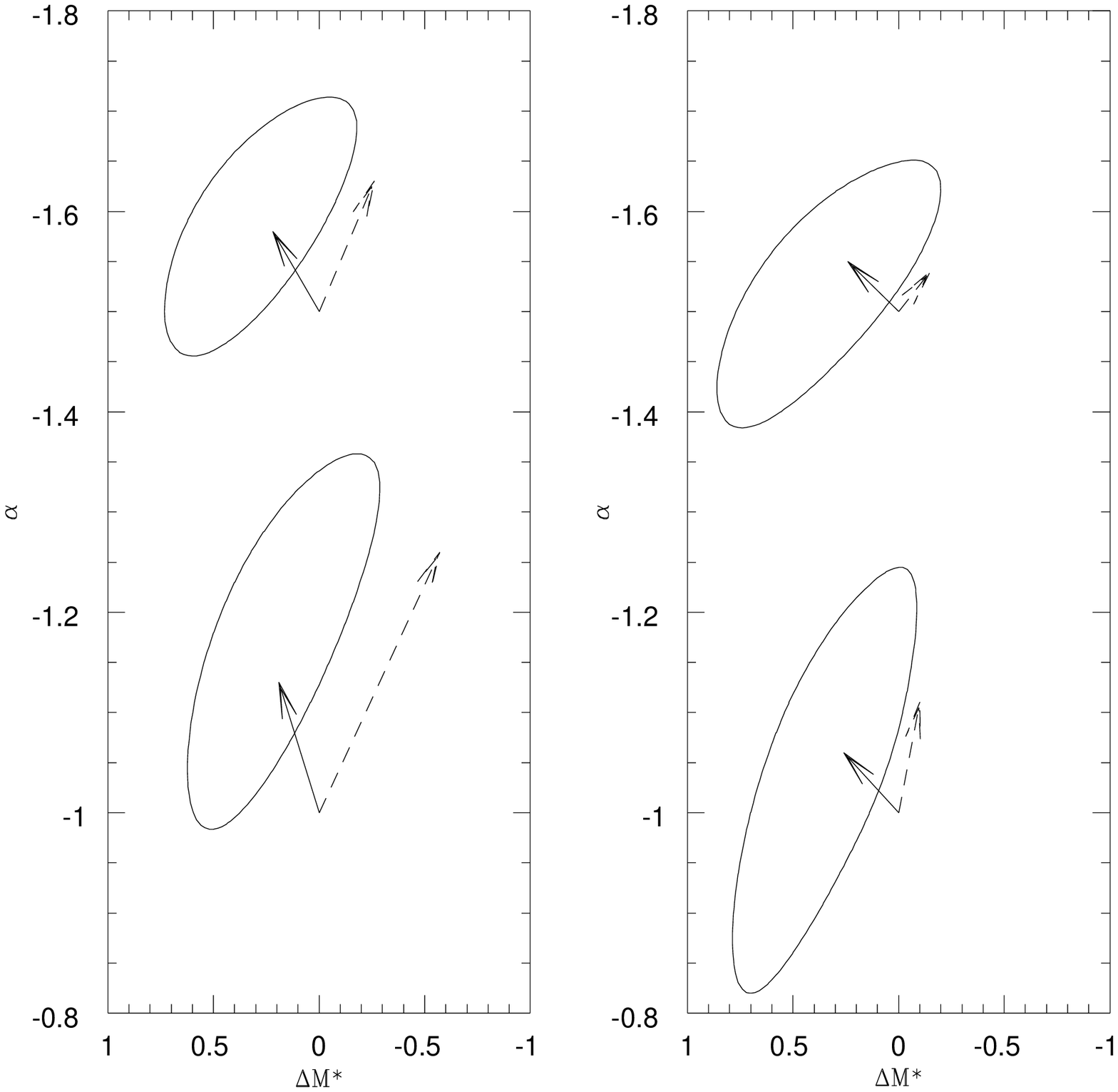]{The finite error in the redshift estimate
results in a bias in the derived
Schecter parameters $M^*$ and $\alpha$ (similar to a Malmquist bias). The figure on the left is for
the errors in redshift given by Eqn 1.  The figure on the right is for
errors in redshift one-third of that, which represent the highest
accuracy obtainable through the photometric redshift method.  The dotted arrow
shows the deviation from the true values of $M^*$ and $\alpha$ when no bias correction is applied.  The solid line shows the deviation in the parameters after we have applied our bias correction.   The 90\%
confidence contours are plotted for the simulations with the bias
correction.  These results were determines for samples of roughly 800 galaxies.}

\figcaption[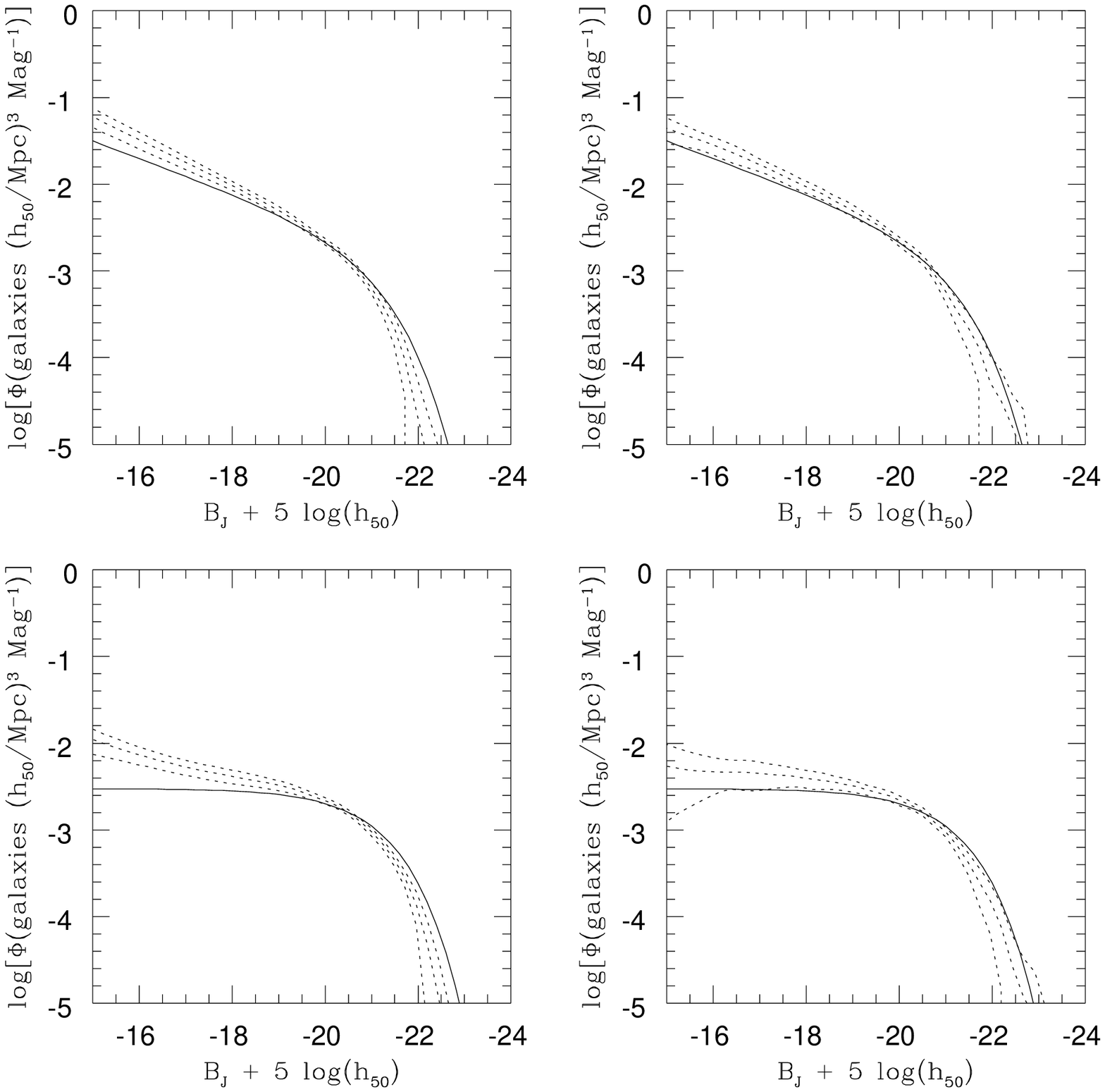]{Results from the simulations.  The two
figures on the top had input luminosity functions of $\Phi^*=2\times
10^{-3},$ $M^*=-21.0,$ $\alpha=-1.5$ and the two on the bottom had
input luminosity functions of $\Phi^*=3\times 10^{-3},$ $M^*=-21.0,$
$\alpha=-1.0$.  The figures on the left had redshift errors given by
equation 1, while the figures on the right are for errors one third of
that, (which represent the highest accuracy obtainable through the
photometric method).  The solid lines are the input luminosity
functions and the three dashed lines represent the recovered
luminosity function and the one sigma spread in the estimates.}

\figcaption[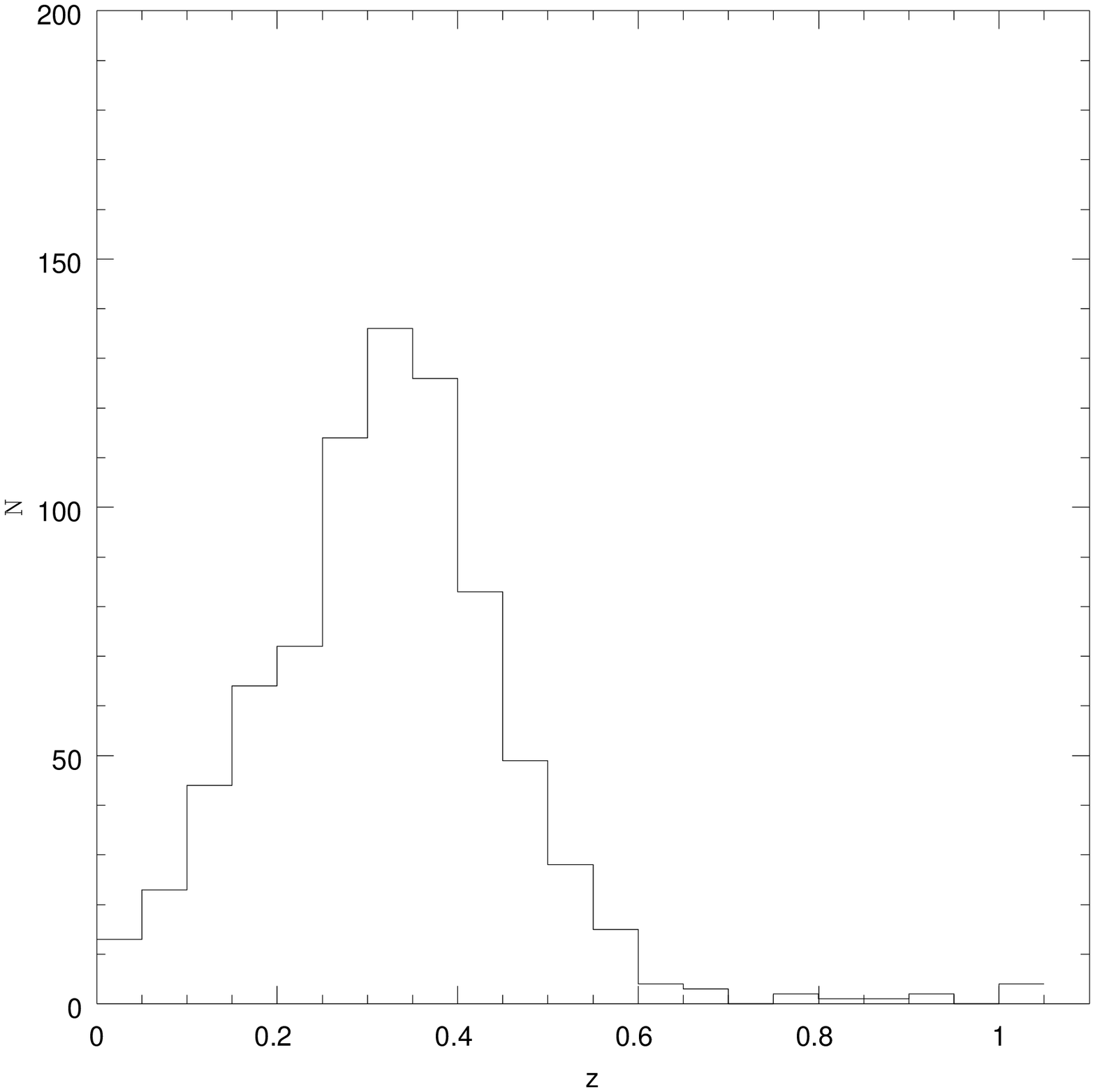]{The redshift distribution for SA68 derived from the
photometric redshift sample}

\figcaption[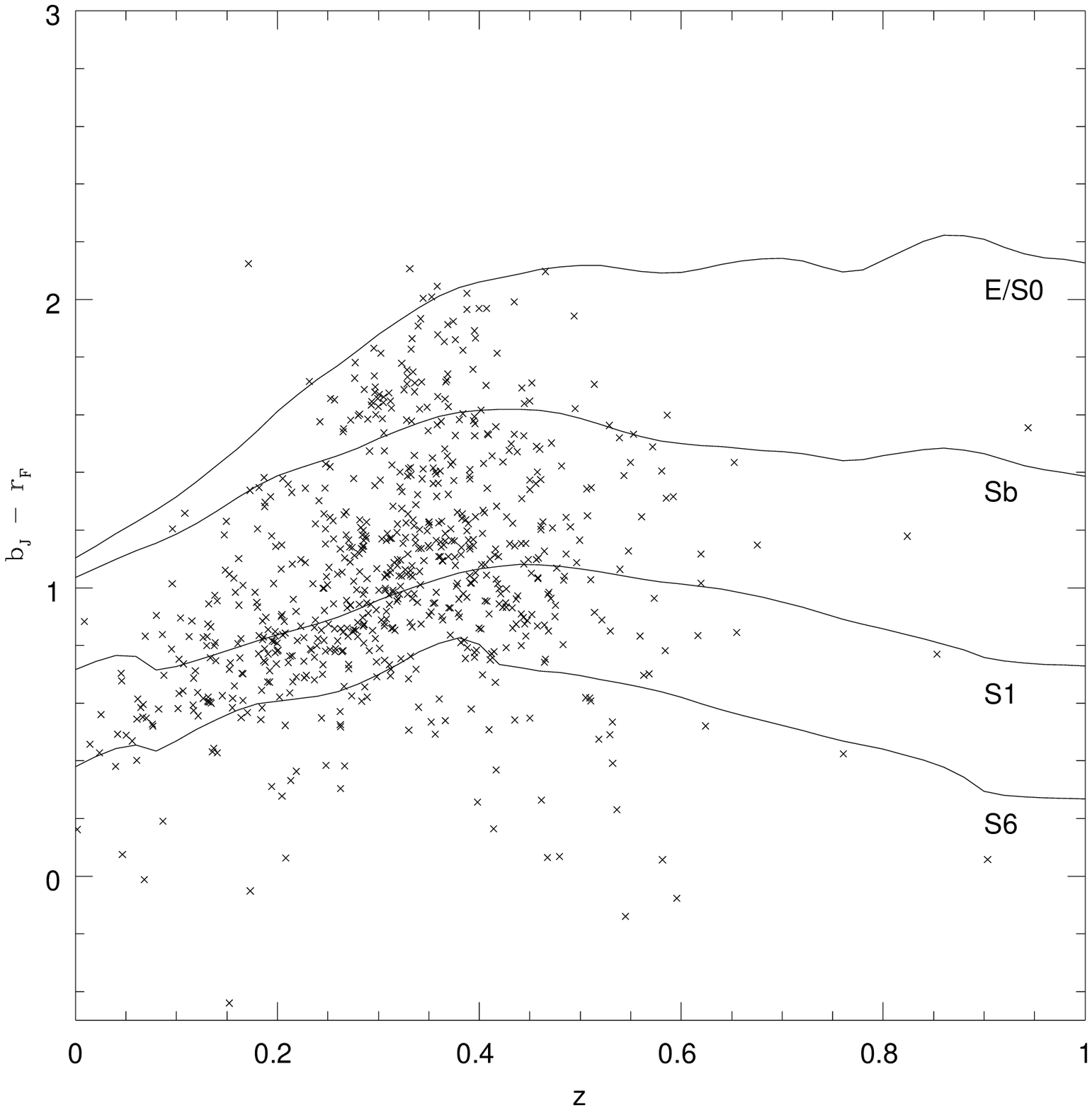]{The colors of the galaxies in our sample
overplotted with the four template spectra from Kinney et al..
K--corrections are interpolated using the two closest template
spectra.}

\figcaption[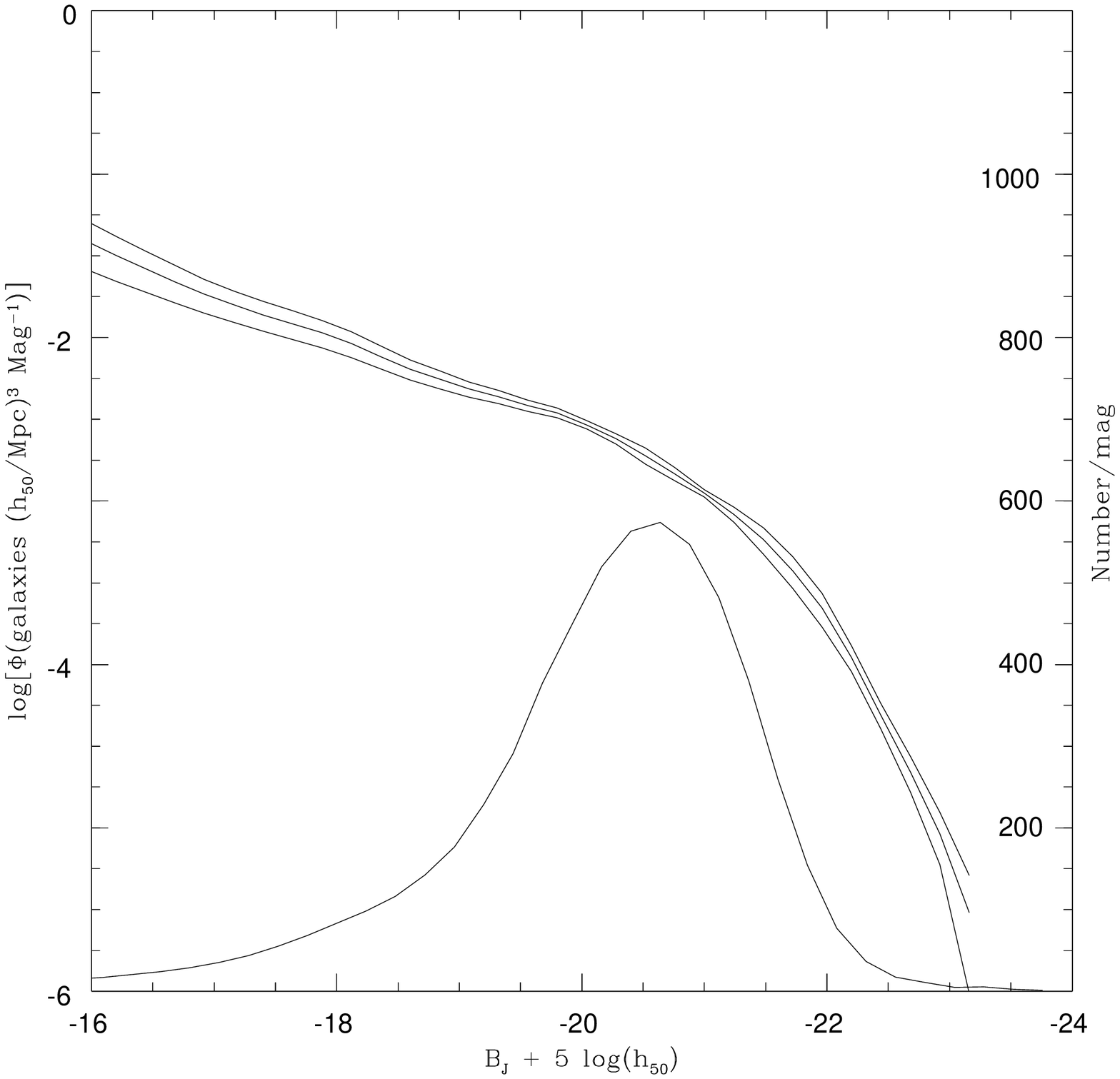]{The resulting `minimal' luminosity
function derived from the photometric redshift sample.  The line is bracketed by its one sigma errors.  The `cut'
luminosity function turns out to be virtually identical except it contains no
estimate fainter that $B_J = -17$.  The distribution in absolute
magnitude is also plotted to give an understanding of how many
galaxies contribute at each magnitude interval.}

\figcaption[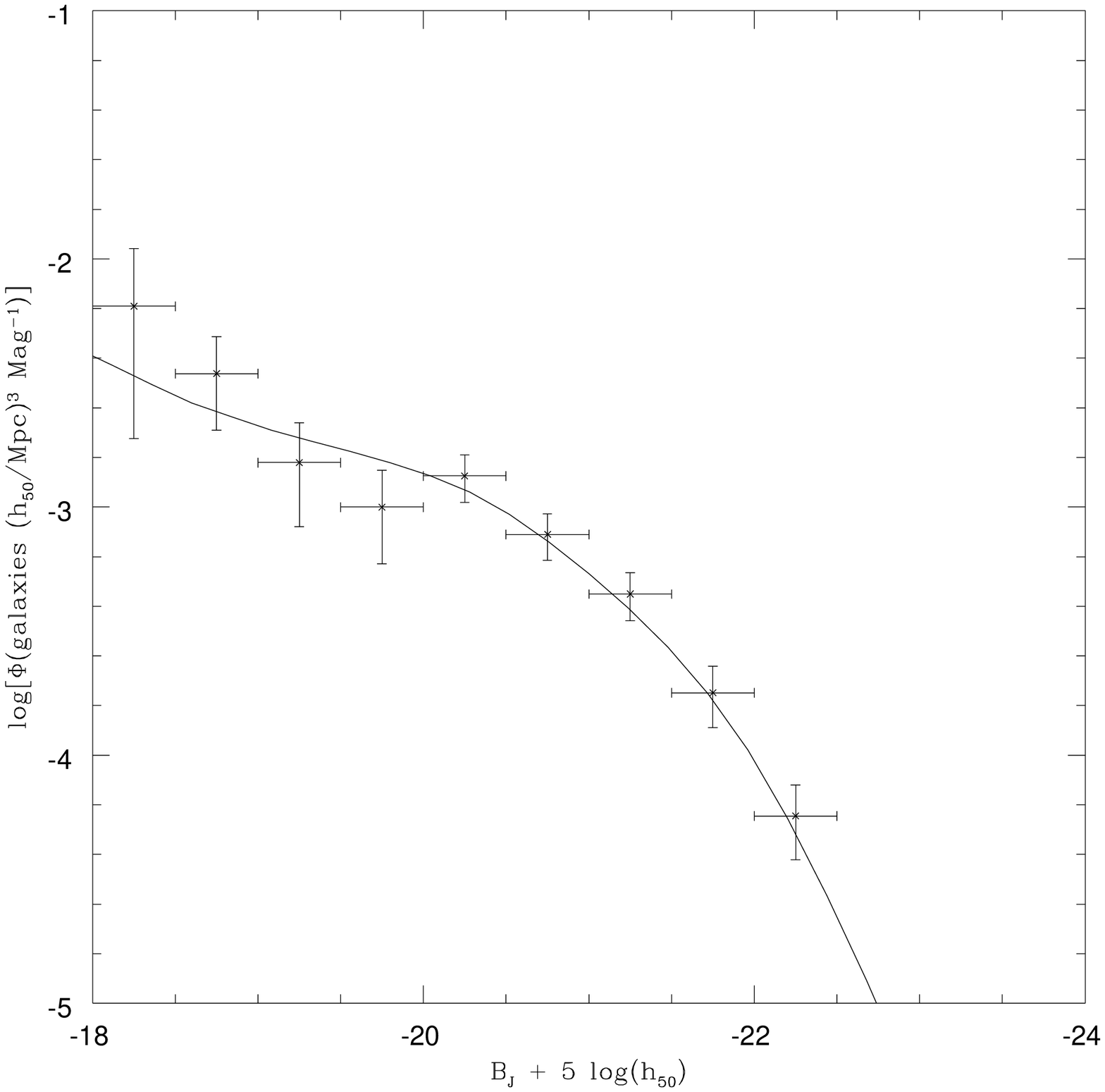]{A comparison of the luminosity function derived from the photometric redshift method ( solid line) with that calculated using the spectroscopic redshifts (points with error bars). To facilitate a direct comparison the photometric sample was cut to similar limits, $b_J=20$,
as the spectroscopic sub-sample.  The two luminosity functions are in good agreement.}

\end{document}